\documentclass[onecolumn,preprintnumbers,amsmath,amssymb,notitlepage]{revtex4}
\usepackage{graphicx}
\usepackage{epsfig}
\usepackage{amsmath}
\usepackage{amsfonts}
\usepackage{amssymb}
\usepackage{textcomp}
\usepackage{feynmf}

\begin{document}

\title{Gauge Theories under Incorporation of a Generalized Uncertainty Principle}

\author{Martin Kober}
\email{kober@fias.uni-frankfurt.de}
\email{kober@th.physik.uni-frankfurt.de}

\affiliation{Frankfurt Institute for Advanced Studies (FIAS),
Institut f\"ur Theoretische Physik, Johann Wolfgang Goethe-Universit\"at, 
Ruth-Moufang-Strasse 1, 60438 Frankfurt am Main, Germany}
\date{\today}

\begin{abstract}
There is considered an extension of gauge theories according to the assumption of a generalized uncertainty principle which
implies a minimal length scale. A modification of the usual uncertainty principle implies an extended shape of matter field 
equations like the Dirac equation. If there is postulated invariance of such a generalized field equation under local gauge 
transformations, the usual covariant derivative containing the gauge potential has to be replaced by a generalized covariant 
derivative. This leads to a generalized interaction between the matter field and the gauge field as well as to 
an additional self-interaction of the gauge field. Since the existence of a minimal length scale seems to be a necessary 
assumption of any consistent quantum theory of gravity, the gauge principle is a constitutive ingredient of the standard 
model, and even gravity can be described as gauge theory of local translations or Lorentz transformations, the presented 
extension of gauge theories appears as a very important consideration.
\end{abstract}

\maketitle

\section{Introduction}

It is widely assumed that a consistent formulation of a quantum theory of gravity implies the existence of a minimal
length scale in nature, which is usually expected to be directly connected to the Planck length. In accordance with this, all
present approaches of a formulation of a quantum theory of gravity contain a minimal length arising in a natural way.
The introduction of a minimal length scale to the description of nature is directly related to the assumption of a generalized
uncertainty principle in quantum mechanics \cite{Maggiore:1993kv},\cite{Kempf:1994su},\cite{Hinrichsen:1995mf}, where the usual
Heisenbergian commutation relation between the position and the momentum operator is extended in such a way that it depends on
the position or on the momentum operator. If it depends on the momentum operator, this implies that there exists not only a
minimal uncertainty between positions and momenta but also of position itself, which represents nothing else than a minimal
length. According to such a generalized uncertainty principle, there arise new representations of the position operator
in momentum space and the momentum operator in position space, respectively. If the generalized momentum operator represented in
position space is inserted to quantum theoretical field equations, this leads to a modification of the dynamics depending on
the scale, which is directly related to the corresponding parameter introduced by the modification of the uncertainty principle.
There have already been considered various implications of such a generalized description of nature to certain aspects of 
quantum mechanics and quantum field theory as they are investigated in
\cite{Kempf:1994su},\cite{Hinrichsen:1995mf},\cite{Kempf:1996ss},\cite{Kempf:1996nk},\cite{Lubo:1999xg},\cite{Chang:2001kn},\cite{Hossenfelder:2003jz},\cite{Harbach:2003qz},\cite{Harbach:2005yu},\cite{Hossenfelder:2004up},\cite{Hossenfelder:2006cw}\cite{Hossenfelder:2007re},\cite{Bang:2006va},\cite{Shibusa:2007ju},\cite{Kim:2008kc}
as well as to many topics related to gravity as they are treated in 
\cite{Maggiore:1993rv},\cite{Maggiore:1993zu},\cite{Scardigli:1999jh},\cite{Capozziello:1999wx},\cite{Crowell:1999up},\cite{Chang:2001bm},\cite{Kim:2006rx},\cite{Park:2007az},\cite{Scardigli:2007bw},\cite{Kim:2007hf},\cite{Bina:2007wj},\cite{Battisti:2007zg},\cite{Zhu:2008cg},\cite{Battisti:2008qi},\cite{Battisti:2008rv},\cite{Vakili:2008zg},\cite{Myung:2009gv},\cite{Myung:2009ur},\cite{Ali:2009zq},\cite{Farmany:2009zz},\cite{Li:2009zz},\cite{Bina:2010ir},\cite{Kim:2010wc}.
A very important topic with respect to the formulation of quantum field theories with a generalized uncertainty principle is
the implication concerning the gauge principle on which all interaction theories of the standard model as well as gravity,
which can be considered to be the gauge theory of local translations or Lorentz rotations, are based on. The issue of gauge
invariance in the context of quantum field theories with a generalized uncertainty principle has already been considered
in \cite{Hossenfelder:2003jz} for example.
In the common description of gauge theories the usual derivative is replaced by a covariant derivative by introducing a gauge
potential which transforms in a specific way to maintain local gauge invariance of a field equation under a certain symmetry
group. Since the position representation of the momentum operator obeying a generalized uncertainty principle contains
additional derivative terms, which also have to be replaced by covariant derivatives to maintain local gauge invariance, there
are included further interaction terms. These additional terms imply a self-interaction of the gauge field even for Abelian
gauge theories like electromagnetism and imply a more complicated interaction between the matter field and the gauge field
leading to extended vertices in the corresponding quantum field theories.
There arises the intricacy that a generalized uncertainty principle leads to an action of infinite order in derivatives
\cite{Hossenfelder:2007re} implying an infinite series of interaction terms. Therefore it has to be considered a series
expansion. In \cite{Hossenfelder:2003jz} a series expansion in the product of the gauge coupling constant
and the squared Planck length has been made. In the present paper an expansion in the modification parameter corresponding to
the squared Planck length will be used instead. This implies that already a calculation to the first order yields additional
interaction terms of the gauge field.
Because of the fact that the gauge principle is, on the one hand, a constitutive ingredient of the standard model and as such
determines the structure of its interactions and, on the other hand, a quantum description of gravity implies the existence of a
minimal length, it seems to be very important to investigate gauge theories under incorporation of a generalized uncertainty
principle. The consideration of gauge theories implies in this context that a generalized uncertainty principle leads
inevitably to a modification not only of the behaviour of free fields but also of the structure of the interactions. Since also
gravity can be considered as gauge theory, the generalized uncertainty principle also has an influence on the dynamics of
gravity. If it is described as gauge theory of local Lorentz rotations, the introduction of the minimal length leads to
generalized classical Einstein field equations and a generalized equivalence principle. This extended gauge theoretic
description of gravity could be seen as a kind of semiclassical approximation to a quantum theory of gravity.

The paper is structured as follows: First, a short repetition of the generalized uncertainty principle, which is formulated in
a way including the time component, is given. Then the modification of the usual formulation of gauge theories as it is used in
the standard model is considered according to a generalized uncertainty principle. After showing how local invariance under a
certain symmetry group can be maintained in the presence of a generalized momentum operator and the corresponding modified free
field equation, the special cases of electromagnetism corresponding to a $U(1)$ gauge theory and gravity corresponding to a
$SO(3,1)$ gauge theory in the chosen formulation are explored. From the extended actions for the gauge field and the matter
field with the additional interaction terms, on the one hand, and the expressions for free quantum fields according to the
corresponding extension of momentum eigenstates, on the other hand, both calculated in an approximation to the first order in
the modification parameter, the Feynman rules of the modified theory of quantum electrodynamics are given. Subsequently,
the consequences for the gauge description of gravity and the corresponding changing of the dynamics of the gravitational field and matter fields in general relativity are investigated.

\section{Generalized Uncertainty Principle and Modified Dirac Equation}

A short repetition of the introduction of a minimal length to quantum theory by postulating a generalized uncertainty principle
between positions and momenta will first be given. A generalized uncertainty principle which implies a minimal length but no
minimal momentum and which does therefore not spoil translation invariance has the following general shape 

\begin{equation}
\left[\hat x^\mu,\hat p_\nu\right]=i\delta^\mu_\nu \left(1+\beta f\left(\hat p\right)\right)
+i\beta g^\mu_\nu\left(\hat p\right),
\end{equation}
where $f\left(\hat p\right)$ and $g^\mu_\nu\left(\hat p\right)$ are general functions depending on the four-momentum and $\beta$ denotes the parameter determining the strength of the modification of the uncertainty principle that is usually assumed
to be connected to the Planck scale. Here is assumed the relativistic case where the time coordinate is included which
leads also to a minimal time scale $\Delta t_0$ corresponding to the minimal length scale $\Delta x_0$.
In the following consideration the special assumption that $f(\hat p)=\hat p^\rho \hat p_\rho$ and $g^\mu_\nu(\hat p)
=2 \hat p^\mu \hat p_\nu$ will be made, meaning that the following special commutation relation between the position
an the momentum operator is valid:

\begin{equation}
\left[\hat x^\mu,\hat p_\nu\right]=i\delta^\mu_\nu \left[1+\beta \hat p^\rho \hat p_\rho\right]+2i\beta \hat p^\mu \hat p_\nu.
\label{generalized_uncertainty_principle}
\end{equation}
This commutation relation corresponds to a generalized uncertainty relation that reads

\begin{equation} 
\Delta x^\mu \Delta p_\mu \geq \frac{1}{2}\left(1+\beta \Delta p^\rho \Delta p_\rho
+\beta \langle p^\rho \rangle \langle p_\rho \rangle\right)+i\left(\beta \Delta p^\mu
\Delta p_\mu+\beta\langle p^\mu \rangle \langle p_\mu \rangle\right)
=\frac{1}{2}\left(1+3\beta \Delta p^\mu \Delta p_\mu
+3\beta \langle p^\rho \rangle \langle p_\rho \rangle\right).
\end{equation}
If one considers the minimal uncertainty for the index of the position operator $\mu$ equal to the index of the momentum
operator $\nu$ and solves the resulting equation for $\Delta p^\mu$, one obtains

\begin{equation}
\Delta p^\mu=\frac{\Delta x^\mu}{3\beta}\pm \sqrt{\left(\frac{\Delta x^\mu}{3\beta}\right)^2-\frac{1}{3\beta}
-\langle p^\rho \rangle \langle p_\rho \rangle}.
\end{equation}
By setting $\Delta p^\mu$ equal to zero the minimal position uncertainty in Minkowski space is obtained

\begin{equation}
l_s=\Delta x^\mu_{min}=\sqrt{3\beta}\sqrt{1+3\beta\langle p^\rho \rangle \langle p_\rho \rangle}
\end{equation}
leading to the smallest length for $\mu=1,...,3$ and the smallest time for $\mu=0$

\begin{equation}
l_s=\sqrt{3\beta}\quad,\quad t_s=\sqrt{3\beta},
\end{equation}
if it is assumed that $\hbar=c=1$ as it is done throughout this paper.
According to the modified uncertainty principle ($\ref{generalized_uncertainty_principle}$), generalized representations of the
position operator in momentum space and the momentum operator in position space have to be given. Since on the right hand side
of ($\ref{generalized_uncertainty_principle}$) there just appears the momentum, the representation in momentum space can be
given directly as

\begin{equation}
\hat p_\mu=p_\mu \quad,\quad \hat x^\mu=i\left(1+\beta p^\rho p_\rho\right)\frac{\partial}{\partial p_{\mu}}
+i2\beta p^\mu p_\rho \frac{\partial}{\partial p_{\rho}}.
\end{equation}
To yield the representation in position space, a series expansion has to be considered, since in this case the momentum
operator becomes nontrivial instead of the position operator and this nontrivial expression depends on the momentum leading to
an iterative definition. To the first order in $\beta$ the operators look as follows:

\begin{equation}
\hat x^\mu=x^\mu \quad,\quad \hat p_\mu=-i\left(1-\beta \partial^{\rho}\partial_\rho\right)\partial_\mu
+\mathcal{O}\left(\beta^2\right).
\label{operators_position-space}
\end{equation}
If the momentum operator in ($\ref{operators_position-space}$) is inserted to the Dirac equation, it reads accordingly

\begin{equation}
\left[i\left(1-\beta \partial^\rho \partial_\rho\right) \gamma^\mu \partial_\mu-m\right]\psi=0.
\label{modified_Dirac_equation}
\end{equation}

\section{The Gauge Principle under Incorporation of a Generalized Uncertainty Principle}

The modification of the formulation of gauge theories according to the generalization of the uncertainty principle
($\ref{generalized_uncertainty_principle}$) and the corresponding modification of a free matter field equation like
the Dirac equation ($\ref{modified_Dirac_equation}$) will now be considered.
It should be emphasized again that in contrast to \cite{Hossenfelder:2003jz}, where is already formulated a Dirac equation and
a field strength tensor containing the generalized uncertainty principle and being gauge invariant, in the present paper an
expansion in the modification parameter will be used implying self-interaction terms of the gauge field in a calculation to the
first order instead of an expansion in the product of the gauge coupling constant and the squared Planck length.
As in the usual case, one starts with the action of the free matter field, which reads in the scenario of this paper as follows:

\begin{equation}
\mathcal{S}=\int d^4 x\ \bar \psi \left[i\left(1-\beta \partial^\rho \partial_\rho\right)\gamma^\mu
\partial_\mu-m\right]\psi
\label{modified_Dirac_action}
\end{equation}
being invariant under certain global symmetry transformations. But in this case there exist additional derivative expressions.
Since the additional derivatives also act on the local unitary transformation operator $U(x)$, they have to be replaced by
covariant derivatives too:

\begin{equation}
\left(\bf{1}-\beta \partial^{\rho}\partial_\rho\right)\partial_\mu 
\rightarrow \left({\bf 1}-\beta D^{\rho}D_\rho\right)D_\mu,
\label{derivatives}
\end{equation}
leading to a transition $\mathcal{S} \rightarrow \mathcal{S}_m$ to the following generalized matter action containing the
coupling of the matter field to the gauge field:   

\begin{equation}
\mathcal{S}_{m}=\int d^4 x\ \bar \psi \left[i\left({\bf 1}-\beta D^\rho D_\rho\right) \gamma^\mu D_\mu-m\right]\psi.
\label{modified_gauge_Dirac_action}
\end{equation}
This action is invariant under global gauge transformations, since the additional term introduced by the generalized 
uncertainty principle transforms as follows:

\begin{eqnarray}
i \beta \bar \psi D^\rho D_\rho \gamma^\mu D_\mu \psi \rightarrow 
i \beta \bar \psi U^{\dagger}(x) U(x) D^\rho U^{\dagger}(x) U(x) D_\rho U^{\dagger}(x) 
\gamma^\mu U(x) D_\mu U^{\dagger}(x) U(x) \psi
=i \beta \bar \psi D^\rho D_\rho \gamma^\mu D_\mu \psi.
\label{symmetry}
\end{eqnarray}
In ($\ref{symmetry}$) it has been used that $U^{\dagger}(x)U(x)={\bf 1}$ and that $U(x)$ either commutes with 
$\gamma^\mu$, if $U(x)$ does not refer to a space-time transformation, or $\gamma^\mu$ transforms according to 
$\gamma^\mu \rightarrow U(x) \gamma^\mu U^{\dagger}(x)$.
Therefore it makes sense to define a generalized covariant derivative in accordance with ($\ref{derivatives}$)

\begin{equation}
\mathcal{D}_\mu \equiv \left({\bf 1}-\beta D^{\rho}D_\rho\right)D_\mu
=\left[{\bf 1}-\beta\left(\partial^{\rho}{\bf 1}+iA^{\rho}\right)\left(\partial_\rho{\bf 1}+iA_\rho \right)\right]
\left(\partial_\mu{\bf 1}+iA_\mu\right),
\label{definition_generalized_covariant_derivative}
\end{equation}
which behaves because of ($\ref{symmetry}$) like the usual covariant derivative under a local gauge transformation

\begin{equation}
\mathcal{D}_\mu \rightarrow U^{\dagger}(x)\mathcal{D}_\mu U(x). 
\label{transformation_generalized_covariant_derivative}
\end{equation}
Accordingly, a generalized field strength tensor for the gauge field also has to be defined, which is done by
replacing the usual covariant derivative by the new covariant derivative

\begin{equation}
\mathcal{F}_{\mu\nu}=-i[\mathcal{D}_\mu, \mathcal{D}_\nu].
\label{definition_generalized_tensor}
\end{equation}
This generalized definition of a field strength tensor is necessary because the action for the gauge field which is built from
($\ref{definition_generalized_tensor}$) has to correspond to a field equation where the generalized uncertainty principle is
contained. It will be shown later that replacing the usual momentum operator by the generalized momentum operator in the free
field equation of a vector field leads, in the approximation where self-interaction terms are neglected, to the same equation
as variation of the action that is built from ($\ref{definition_generalized_tensor}$).
Of course, the generalized field strength tensor is still invariant under local gauge transformations. 
The generalization of the gauge principle according to 
($\ref{modified_Dirac_action}$),($\ref{derivatives}$),($\ref{modified_gauge_Dirac_action}$),($\ref{symmetry}$),($\ref{definition_generalized_covariant_derivative}$),($\ref{transformation_generalized_covariant_derivative}$),($\ref{definition_generalized_tensor}$) holds for all gauge theories independent of the special symmetry group that is
considered. In the special case of a Yang-Mills theory there can be built the following generalized action of the gauge field
$\mathcal{S}_{g}=\frac{1}{4}\int d^4 x\ {\text tr}\left[\mathcal{F}_{\mu\nu} \mathcal{F}^{\mu\nu}\right]$ 
by using ($\ref{definition_generalized_tensor}$) and thus the complete action of the modified Yang-Mills gauge theory reads

\begin{equation}
\mathcal{S}_{c}=\int d^4 x \left(\bar\psi \left(i\gamma^\mu
\mathcal{D}_\mu-m\right)\psi-\frac{1}{4}{\text tr}\left[\mathcal{F}_{\mu\nu}
\mathcal{F}^{\mu\nu}\right]\right).
\label{generalized_action}
\end{equation}
Thus there has been obtained a generalized action for a Yang-Mills gauge theory in the presence of a generalized uncertainty
principle. This action will be explored explicitly for the simplest case of the $U(1)$ gauge group in the next section, and
the implications for the corresponding theory of quantum electrodynamics will be considered.

\section{Modified Quantum Electrodynamics}

In this section the extended quantum field theoretic description arising from the generalization of gauge theories according to
the generalization of the uncertainty principle will be considered as it has been described above. Specifically,
the special case of the $U(1)$ gauge group corresponding to quantum electrodynamics will be regarded. The actions for the gauge
field and the matter field in terms of the gauge potential $A_\mu$ will be calculated and the corresponding Feynman rules will
be determined. Similar considerations referring to the other series expansion can be found in \cite{Hossenfelder:2004up}.

\subsection{Calculation of the Generalized Action}

If the matter sector ($\ref{modified_gauge_Dirac_action}$) of the generalized action ($\ref{generalized_action}$)
is now calculated to the first order in the modification parameter $\beta$ for the special case of a $U(1)$ gauge 
group, one obtains 

\begin{eqnarray}
\mathcal{S}_{m}&=&\int d^4 x\ \bar \psi\left(i\gamma^\mu \mathcal{D}_\mu-m \right)\psi
=\int d^4 x\ \bar \psi \left[i\left(1-\beta D^\rho D_\rho\right) \gamma^\mu D_\mu-m\right]\psi\nonumber\\
&=&\int d^4 x\ \bar \psi\left\{i\gamma^\mu\left[1-\beta \left(\partial^{\rho}+iA^{\rho}\right)\left(\partial_\rho+iA_\rho\right)\right]
\left(\partial_\mu+iA_\mu\right)-m\right\}\psi\nonumber\\
&=&\int d^4 x\ \bar \psi\left\{i\gamma^\mu\left[\partial_\mu+iA_\mu-\beta \left(\partial^{\rho}\partial_{\rho}\partial_\mu+i\partial^\rho A_\rho \partial_\mu
+2iA_{\rho}\partial^\rho \partial_\mu
-A^\rho A_\rho \partial_\mu
+i\partial^\rho \partial_\rho A_\mu\right.\right.\right.\nonumber\\
&&\left.\left.\left.+2i\partial_\rho A_\mu \partial^\rho
+iA_\mu \partial^\rho \partial_\rho-\partial^\rho A_\rho A_\mu-2A_\rho \partial^\rho A_\mu
-2A_\rho A_\mu \partial^\rho
-iA^\rho A_\rho A_\mu\right)\right]-m\right\}\psi.
\label{matter_action}
\end{eqnarray}
This action, of course, describes a much more complicated interaction structure between the matter field and the gauge field
than the usual action giving rise to new vertices where more than one gauge boson interact with the matter field. If the 
gauge sector is to be calculated explicitly, the shape of the generalized field strength tensor $\mathcal{F}_{\mu\nu}$ defined in ($\ref{definition_generalized_tensor}$) first has to be calculated. For the case of electromagnetism, inserting
($\ref{definition_generalized_covariant_derivative}$) to ($\ref{definition_generalized_tensor}$) yields

\begin{eqnarray}
\mathcal{F}_{\mu\nu}&=&-i\left[\mathcal{D}_\mu,\mathcal{D}_\nu\right]
=-i\left[\left(1-\beta D^\rho D_\rho\right)D_\mu,\left(1-\beta D^\rho D_\rho\right)D_\nu\right]\nonumber\\
&=&-i\left\{\left(1-\beta D^\rho D_\rho\right)^2\left[D_\mu,D_\nu\right]
+\left(1-\beta D^\rho D_\rho\right)\left[D_\mu,\left(1-\beta D^\rho D_\rho\right) \right]D_\nu\right.\nonumber\\
&&\left.+\left(1-\beta D^\rho D_\rho\right)\left[\left(1-\beta D^\rho D_\rho\right),D_\nu\right]D_\mu
+\left[\left(1-\beta D^\rho D_\rho\right),\left(1-\beta D^\rho D_\rho\right)\right]D_\mu D_\nu\right\}\nonumber\\
&=&-i\left\{\left(1-\beta D^\rho D_\rho\right)^2\left[D_\mu,D_\nu\right]
-\left(1-\beta D^\rho D_\rho\right)\beta\left[D_\mu,D^\rho D_\rho\right]D_\nu
-\left(1-\beta D^\rho D_\rho\right)\beta\left[D^\rho D_\rho,D_\nu\right]D_\mu\right\}\nonumber\\
&=&-i\left\{\left(1-\beta D^\rho D_\rho\right)^2 \left[D_\mu, D_\nu\right]
-\left(1-\beta D^\rho D_\rho\right)\beta D^\rho \left[D_\mu,D_\rho\right]D_\nu
-\left(1-\beta D^\rho D_\rho\right)\beta \left[D_\mu,D^\rho\right]D_\rho D_\nu\right.\nonumber\\
&&\left.+\left(1-\beta D^\rho D_\rho\right)\beta D^\rho \left[D_\nu,D_\rho\right]D_\mu
+\left(1-\beta D^\rho D_\rho\right)\beta \left[D_\nu,D^\rho\right]D_\rho D_\mu\right\}\nonumber\\
&=&\left(1-\beta D^\rho D_\rho\right)\left[\left(1-\beta D^\rho D_\rho\right)F_{\mu\nu}
-\beta\left(D^\rho F_{\mu\rho}D_\nu-D^\rho F_{\nu\rho}D_\mu\right)
-\beta \left(F_{\mu\rho}D^{\rho}D_\nu-F_{\nu\rho}D^{\rho}D_\mu\right)\right].
\end{eqnarray}
This means that the generalized field strength tensor $\mathcal{F}_{\mu\nu}$ expressed by the usual field strength tensor 
$F_{\mu\nu}$ and the usual covariant derivative $D_\mu$ to the first order in $\beta$ reads

\begin{equation}
\mathcal{F}_{\mu\nu}=F_{\mu\nu}-2\beta D^\rho D_\rho F_{\mu\nu}
-\beta\left(D^\rho F_{\mu\rho}D_\nu-D^\rho F_{\nu\rho}D_\mu\right)
-\beta \left(F_{\mu\rho}D^{\rho}D_\nu-F_{\nu\rho}D^{\rho}D_\mu\right)+\mathcal{O}\left(\beta^2\right).
\label{generalized_tensor}
\end{equation}
The corresponding action of the gauge field sector reads accordingly

\begin{eqnarray}
\mathcal{S}_{g}&=&\frac{1}{4}\int d^4 x {\mathcal{F}}_{\mu\nu}{\mathcal{F}}^{\mu\nu}\nonumber\\
&=&\frac{1}{4}\int d^4 x \left\{F_{\mu\nu}F^{\mu\nu}-\beta\left[2 F_{\mu\nu}\left(D^\rho D_\rho F^{\mu\nu}\right)+
F_{\mu\nu}\left(F^{\mu\rho}D_{\rho}D^\nu-F^{\nu\rho}D_{\rho}D^\mu\right)+F_{\mu\nu}\left(D_\rho F^{\mu\rho}D^\nu
-D_\rho F^{\nu\rho}D^\mu\right)\right.\right.\nonumber\\
&&\left.\left.+2\left(D^\rho D_\rho F_{\mu\nu}\right)F^{\mu\nu}+\left(F_{\mu\rho}D^{\rho}D_\nu-F_{\nu\rho}D^{\rho}D_\mu\right)F^{\mu\nu}
+\left(D^\rho F_{\mu\rho}D_\nu-D^\rho F_{\nu\rho}D_\mu\right)F^{\mu\nu}\right]\right\}+\mathcal{O}\left(\beta^2\right)\nonumber\\
&=&\frac{1}{4}\int d^4 x \left[F_{\mu\nu}F^{\mu\nu}-4\beta F_{\mu\nu}D^\rho D_\rho F^{\mu\nu}-4\beta 
F_{\mu\nu}F^{\mu\rho}D_{\rho}D^\nu-4\beta F_{\mu\nu}D_\rho F^{\mu\rho}D^\nu 
\right]+\mathcal{O}\left(\beta^2\right).\nonumber\\
\end{eqnarray}
Inserting the expression for the usual covariant derivative yields for the generalized action of the gauge field to the first
order in $\beta$

\begin{eqnarray}
\mathcal{S}_{g}&=&\frac{1}{4}\int d^4 x
\left[F_{\mu\nu}F^{\mu\nu}
-\beta\left(4 F_{\mu\nu} \partial_\rho \partial^\rho F^{\mu\nu}
+4i\partial_\rho A^\rho F_{\mu\nu}F^{\mu\nu}
+8i F_{\mu\nu}A_\rho \partial^\rho F^{\mu\nu}
\right.\right.\nonumber\\&&\left.\left.
+8i\partial^\rho A_\nu F_{\mu\rho}F^{\mu\nu}
+4i A^\rho F_{\mu\rho} \partial_\nu F^{\mu\nu}
-8 A^\rho F_{\mu\rho} A_\nu F^{\mu\nu}
-4 A^{\rho}A_\rho F_{\mu\nu}F^{\mu\nu}\right)\right]
\end{eqnarray}
and expressing the field strength tensor by the gauge potential finally leads to 

\begin{eqnarray}
\mathcal{S}_{g}&=&\frac{1}{4}\int d^4 x
\left[2\partial_\mu A_\nu \partial^\mu A^\nu
-2\partial_\mu A_\nu \partial^\nu A^\mu+\beta\left(
-8\partial_\mu A_\nu \partial_\rho \partial^\rho \partial^\mu A^\nu
+8\partial_\mu A_\nu \partial_\rho \partial^\rho \partial^\nu A^\mu
\right.\right.\nonumber\\&&\left.\left.
-8i\partial_\rho A^\rho \partial_\mu A_\nu \partial^\mu A^\nu
+8i\partial_\rho A^\rho \partial_\mu A_\nu \partial^\nu A^\mu
-16i\partial_\mu A_\nu A_\rho \partial^\rho \partial^\mu A^\nu
+16i\partial_\mu A_\nu A_\rho \partial^\rho \partial^\nu A^\mu
\right.\right.\nonumber\\&&\left.\left.
-8i\partial^\rho A_\nu \partial_\mu A_\rho \partial^\mu A^\nu
+8i\partial^\rho A_\nu \partial_\mu A_\rho \partial^\nu A^\mu
+8i\partial^\rho A_\nu \partial_\rho A_\mu \partial^\mu A^\nu
-8i\partial^\rho A_\nu \partial_\rho A_\mu \partial^\nu A^\mu
\right.\right.\nonumber\\&&\left.\left.
-4i A^\rho \partial_\mu A_\rho \partial_\nu \partial^\mu A^\nu
+4i A^\rho \partial_\mu A_\rho \partial_\nu \partial^\nu A^\mu
+4i A^\rho \partial_\rho A_\mu \partial_\nu \partial^\mu A^\nu
-4i A^\rho \partial_\rho A_\mu \partial_\nu \partial^\nu A^\mu
\right.\right.\nonumber\\&&\left.\left.
+8 A^\rho \partial_\mu A_\rho A_\nu \partial^\mu A^\nu
-8 A^\rho \partial_\mu A_\rho A_\nu \partial^\nu A^\mu
-8 A^\rho \partial_\rho A_\mu A_\nu \partial^\mu A^\nu
+8 A^\rho \partial_\rho A_\mu A_\nu \partial^\nu A^\mu
\right.\right.\nonumber\\&&\left.\left.
+8 A^{\rho}A_\rho \partial_\mu A_\nu \partial^\mu A^\nu
-8 A^{\rho}A_\rho \partial_\mu A_\nu \partial^\nu A^\mu\right)\right].
\label{gauge_field_action}
\end{eqnarray}
If just the terms without self-interaction are considered, one obtains as field equation for the free 
electromagnetic potential 

\begin{equation}
\partial_\mu F^{\mu\nu}-2\beta\partial_\rho \partial^\rho \partial_\mu F^{\mu\nu}=0.
\end{equation}
For Lorentz gauge $\partial_\mu A^\mu=0$ this means

\begin{equation}
\partial_\mu \partial^\mu A^\nu-2\beta\partial_\rho \partial^\rho \partial_\mu \partial^\mu A^\nu=0.
\end{equation}
This is exactly the equation that is obtained if the wave equation of a massless vector field $\partial_\mu \partial^\mu
A^\nu=0$ is generalized according to ($\ref{generalized_uncertainty_principle}$) and ($\ref{operators_position-space}$),
which means that the definition of the generalized field strength tensor yields the correct free field limit. 

\subsection{Propagators}

In the usual setting of quantum field theory, fields are expanded in terms of momentum eigenstates
corresponding to plane waves. If the fields are quantized, the coefficients of plane wave modes become
operators that play the role of creation and annihilation operators of particles. Because of the modified commutation relation 
between position and momentum operators ($\ref{generalized_uncertainty_principle}$), the representation of the momentum operator in position space is changed ($\ref{operators_position-space}$) and therefore also the shape of the momentum eigenstates is modified, which leads to a different expression for a quantum field. The modified momentum eigenstates $|p\rangle$ fulfill the equation

\begin{equation}
\hat p_\mu |p\rangle=p_\mu |p\rangle
\end{equation}
where $p_\mu$ describes the eigenvalue to the eigenstate $|p\rangle$. The momentum eigenstates have the following shape:

\begin{equation}
|p\rangle=\exp\left[i\left(1-\beta p^\rho p_\rho\right)p_\mu x^\mu\right]+\mathcal{O}\left(\exp\left(\beta^2\right)\right),
\end{equation}
if they are expressed to the first order in $\beta$, since if there are only taken terms to the first order one obtains

\begin{equation}
\hat p_\mu \exp\left[i\left(1-\beta p^\rho p_\rho\right)p_\nu x^\nu\right]
=-i\left(1-\beta\partial^\sigma \partial_\sigma\right)\partial_\mu
\exp\left[i\left(1-\beta p^\rho p_\rho\right)p_\nu x^\nu\right]
=p_\mu \exp\left[i\left(1-\beta p^\rho p_\rho\right)p_\nu x^\nu\right].
\end{equation}
If one now considers quantum field theory, the quantum fields have to be expanded by using these 
generalized momentum eigenstates. It will be considered the case of a scalar field, which can be assigned to
the cases of a spinor field and a vector field as they appear in quantum electrodynamics easily at the end   

\begin{equation}
\phi({\bf x},t)=\int \frac{d^3 p}{\sqrt{\left(2\pi\right)^3 2p_0}}\left[a({\bf p})e^{i\left(1-\beta p^\rho p_\rho\right)p_\mu x^\mu}+a^{\dagger}({\bf p})e^{-i\left(1-\beta p^\rho p_\rho\right)p_\mu x^\mu}\right].
\label{modified_quantum_field}
\end{equation}
The commutation relations for the creation and annihilation operators have to remain the same since the Fock space 
structure referring to many particles is not changed by the generalized uncertainty relation which means
$\left[a({\bf p}),a^{\dagger}({\bf p^{\prime}})\right]=\delta^3({\bf p}-{\bf p^{\prime}})$.
The propagator is now obtained as usual by building the expectation value of the time ordered product of the generalized 
expression of the quantum field (\ref{modified_quantum_field}) at two different space-time points with respect to the 
vacuum state $|0\rangle$ of the quantum field

\begin{equation}
G(x-y)=\langle 0|T\left\{\phi(x)\phi(y)\right\}|0\rangle
\label{operator_product}
\end{equation}
where $T$ denotes the time ordering operator.
If one uses ($\ref{modified_quantum_field}$) in ($\ref{operator_product}$), one obtains the expression for the 
propagator in position space

\begin{equation}
G(x-y)=i\int \frac{d^4 p}{(2\pi)^4}\frac{e^{i\left(1-\beta p^2\right)p\left(x-y\right)}}{p^2-m^2+i\epsilon}.
\end{equation}
One has to built the inner product with the momentum eigenstates to obtain the corresponding propagator in momentum space 

\begin{equation}
G(p)=i\int \frac{d^4 z}{(2\pi)^4}\frac{d^4 p^{\prime}}{(2\pi)^4}\frac{e^{i\left(1-\beta p^{\prime 2}
\right)p^{\prime}z-i\left(1-\beta p^2\right)pz}}{p^{\prime 2}-m^2+i\epsilon}.
\end{equation}
To perform the integral it is useful to define $k_\mu$ as $k_\mu=\left(1-\beta p^\rho p_\rho\right)p_\mu$ and to substitute
$p_\mu$ by $k_\mu$ leading to

\begin{eqnarray}
G(k)&=&i\int \frac{d^4 z}{(2\pi)^4}\frac{d^4 k^{\prime}}{(2\pi)^4 J\left(p\left(k^{\prime}\right)\right)}\frac{e^{ik^{\prime}z-ikz}}
{\left(p^2(k^{\prime})-m^2+i\epsilon\right)}
=i\int \frac{d^4 k^{\prime}}{(2\pi)^4
J\left(p\left(k^{\prime}\right)\right)}\frac{\delta \left(k^{\prime}-k\right)}
{\left(p^2(k^{\prime})-m^2+i\epsilon\right)}\nonumber\\
&=&\frac{1}{J\left(p\left(k\right)\right)}\frac{i}{\left(p^{2}(k)-m^2+i\epsilon\right)}
\end{eqnarray}
where $J(p)$ has been defined as the Jacobian determinant of the transformation between $k_\mu$ and $p_\mu$

\begin{equation}
J(p)=\det \left[\frac{\partial k_\mu}{\partial p_\nu}\right]=\left(1-\beta p^\rho p_\rho\right)^4-2\beta\left(1-\beta p^\rho p_\rho\right)^3 p^\sigma p_\sigma+4\beta^2\left(1-\beta p^\rho p_\rho\right)^2\left(-p_0^2 p_1^2+p_1^2 p_2^2+p_2^2 p_3^2-p_3^2 p_0^2\right).
\end{equation}
Resubstituting $k_\mu$ with $p_\mu$ yields the propagator in momentum space

\begin{equation}
i\Delta(p)=G(p)=\frac{1}{J\left(p\right)}\frac{i}{p^{2}-m^2+i\epsilon}.
\label{generalized_propagator}
\end{equation}
To obtain the corresponding propagator for a spinor field and a vector field, the expressions which are analogue to ($\ref{modified_quantum_field}$) are considered

\begin{eqnarray}
\psi({\bf x},t)&=&\sum_{\pm s} \int\frac{d^3 p}{\left(2\pi\right)^{\frac{3}{2}}}\sqrt{\frac{m}{p_0}}\left[b({\bf p},s) u({\bf p},s)e^{i\left(1-\beta p^\rho p_\rho\right)p_\mu x^\mu}+d^{\dagger}({\bf p},s) v({\bf p},s) e^{-i\left(1-\beta p^\rho p_\rho\right)p_\mu x^\mu}\right],\nonumber\\
\psi^{\dagger}({\bf x},t)&=&\sum_{\pm s}\int\frac{d^3p}{\left(2\pi\right)^{\frac{3}{2}}}\sqrt{\frac{m}{p_0}}
\left[b^{\dagger}({\bf p},s) u^{\dagger}({\bf p},s)e^{-i\left(1-\beta p^\rho p_\rho\right)p_\mu x^\mu}+d({\bf p},s) v^{\dagger}({\bf p},s) e^{i\left(1-\beta p^\rho p_\rho\right)p_\mu x^\mu}\right]
\end{eqnarray}
with $\{b({\bf p},s),b^{\dagger}({\bf p^{\prime}},s^{\prime})\}=\delta^3({\bf p}-{\bf p^{\prime}})\delta_{s s^{\prime}}$, 
$\{d({\bf p},s),d^{\dagger}({\bf p^{\prime}},s^{\prime})\}=\delta^3({\bf p}-{\bf p^{\prime}})\delta_{s s^{\prime}}$ and

\begin{equation}
A_\mu({\bf x},t)=\sum_{\lambda=1}^2 \int \frac{d^3 p}{\sqrt{\left(2\pi\right)^3 2p_0}}
\left[a({\bf p},\lambda)\epsilon_\mu({\bf p},\lambda)e^{i\left(1-\beta p^\rho p_\rho\right)p_\mu x^\mu}
+a^{\dagger}({\bf p},\lambda)\epsilon_\mu({\bf p},\lambda)e^{-i\left(1-\beta p^\rho p_\rho\right)p_\mu x^\mu}\right]
\end{equation}
with $\left[a({\bf p},\lambda),a^{\dagger}({\bf p^{\prime}},\lambda^{\prime})\right]=\delta^3({\bf p}-{\bf
p^{\prime}})\delta_{\lambda\lambda^{\prime}}$. 
Besides there are needed as usual the following relations for the summation over the spin states: 
 
\begin{equation}
\sum_{\pm s} u_a({\bf p},s) \bar u_b({\bf p},s)=\left(\frac{p\!\!\!/+m}{2m}\right)_{ab}\quad,\quad 
\sum_{\pm s} v_a({\bf p},s) \bar v_b({\bf p},s)=\left(\frac{p\!\!\!/-m}{2m}\right)_{ab}\quad,\quad
\sum_{\lambda=1}^2 \epsilon_\mu({\bf p},\lambda)\epsilon_\nu({\bf p},\lambda)=-g_{\mu\nu},
\end{equation}
where the last equation referring to the polarization vector of the electromagnetic field only 
holds in Feynman gauge.\\
A calculation that is analogue to the derivation of the propagator of the scalar field 
($\ref{generalized_propagator}$) yields for the propagator of a spinor field describing an electron

\begin{equation}
i\Delta_{ab}(p)=\frac{1}{J\left(p\right)}\left(\frac{i}{p\!\!\!/-m+i\epsilon}\right)_{ab}
\end{equation} 

\begin{fmffile}{gQED1}
\begin{fmfgraph}(40,25)
 \fmfleft{l1}
 \fmf{fermion}{l1,l2} 
 \fmfright{l2}
\end{fmfgraph}
\end{fmffile}
\\ \noindent
and for the propagator of a vector field describing a photon

\begin{equation}
i\Delta_{\mu\nu}(p)=\frac{1}{J\left(p\right)
}\frac{-ig_{\mu\nu}}{p^2+i\epsilon},
\end{equation}

\begin{fmffile}{gQED2}
\begin{fmfgraph}(40,25)
 \fmfleft{l1}
 \fmf{boson}{l1,l2}
 \fmfright{l2}
\end{fmfgraph}
\end{fmffile}
\\ \noindent 
if Feynman gauge is used as it is presupposed. 

\subsection{Vertices}

To extract the vertices arising from the generalized gauge theory of quantum electrodynamics, the interaction terms within the actions ($\ref{matter_action}$) and ($\ref{gauge_field_action}$) have to be considered. Because of the additional 
terms there arise additional vertices where two and three photons interact with the matter field and there also arise 
vertices between photons, since the generalization of gauge theories causes self-interaction terms of the gauge field 
even in the Abelian case. In the following notation $p_\mu$ denotes the four-momentum of a fermion running into the vertex
and the $k_\mu$ denote the four-momenta of the bosons, if they appear in the expression of the respective vertex. Since the 
bosons are indistinguishable, any of the $k_\mu$ can be identified with any boson.\\
The vertex of the interaction between the matter field and the electromagnetic field referring to one photon to the 
first order in $\beta$ reads

\begin{eqnarray}
v^\lambda=-\gamma^\mu\left[\delta^\lambda_\mu+\beta\left(k^{\lambda}p_{\mu}
+2p^\lambda p_{\mu}\right.\right.
\left.\left.+\delta^{\lambda}_\mu k^\rho k_{\rho}+2\delta^{\lambda}_\mu k^{\rho}
p_{\rho}+\delta^\lambda_\mu p^\rho p_{\rho}\right)\right].
\end{eqnarray}

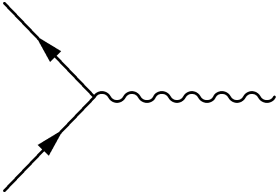
\setlength{\unitlength}{1mm}
\unitlength = 1mm
\begin{fmffile}{gQED3}
 \begin{fmfgraph}(40,25)
  \fmfleft{l1,l2}
  \fmfright{r1}  
  \fmf{fermion}{l1,v1,l2}
  \fmf{boson}{v1,r1} 
 \end{fmfgraph}
\end{fmffile}
\\ \noindent
In this extended case from ($\ref{matter_action}$) it can be seen that there also exists a vertex where two photons interact
with the matter particle. This vertex reads 

\begin{eqnarray}
v^{\lambda\rho}=-\beta\gamma^\mu\left(\delta^{\lambda\rho}p_\mu+\delta^\lambda_\mu k_1^\rho+2\delta^\lambda_\mu k_2^\rho
+2\delta^\lambda_\mu p^\rho\right).
\end{eqnarray}

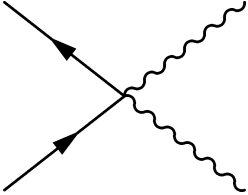
\begin{fmffile}{gQED4}
 \begin{fmfgraph}(40,25)
  \fmfleft{l1,l2} 
  \fmfright{r1,r2} 
  \fmf{fermion}{l1,v1,l2}
  \fmf{boson}{r1,v1} 
  \fmf{boson}{v1,r2}
 \end{fmfgraph}
\end{fmffile}
\\ \noindent
Finally there also exists a vertex with three photons interacting with the matter particle looking as follows:

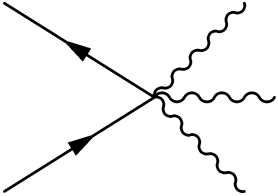
\begin{eqnarray}
v^{\mu\lambda\rho}=-\beta \gamma^\mu \delta^{\lambda\rho}.
\end{eqnarray}

\begin{fmffile}{gQED5}  
 \begin{fmfgraph}(40,25)
  \fmfleft{l1,l2}
  \fmfright{r1,r2,r3}
  \fmf{fermion}{l1,v1,l2}
  \fmf{boson}{v1,r1}
  \fmf{boson}{v1,r2}
  \fmf{boson}{v1,r3}
 \end{fmfgraph}
\end{fmffile}
\\ \noindent
There are two vertices corresponding to a self-interaction of the electromagnetic field. One vertex describes 
the interaction of three photons with each other looking as follows:

\begin{eqnarray}
v^{\mu\nu\lambda}=
\beta\left(2 k_1^{\mu} k_{2\rho} k_3^{\rho} \delta^{\nu\lambda}
-2 k_1^{\mu} k_{2}^{\lambda} k_3^{\nu} 
+4 k_{1\rho} k_{3}^{\rho} k_3^{\nu} \delta^{\mu\lambda}
-4 k_1^{\lambda} k_{3}^{\nu} k_3^{\mu} 
\right.\nonumber\\\left.
+2 k_1^{\nu} k_{2\rho} k_3^{\rho} \delta^{\mu\lambda}
-2 k_1^{\nu} k_{2}^{\lambda} k_3^{\mu} 
-2 k_{1\rho} k_{2}^{\rho} k_3^{\nu} \delta^{\mu\lambda}
+2 k_{1\rho} k_{2}^{\rho} k_3^{\mu} \delta^{\nu\lambda}
\right.\nonumber\\\left.
+ k_{2\rho} k_{3}^{\rho} k_3^{\lambda} \delta^{\mu\nu}
- k_2^{\lambda} k_{3\rho} k_{3}^{\rho} \delta^{\mu\nu}
- k_2^{\mu} k_{3}^{\nu} k_3^{\lambda} 
+ k_{2}^{\mu} k_{3\rho} k_3^{\rho} \delta^{\nu\lambda}\right)
\end{eqnarray}

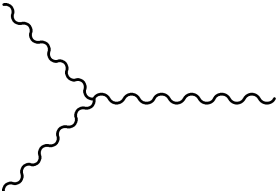
\begin{fmffile}{gQED6}
 \begin{fmfgraph}(40,25)
  \fmfleft{l1,l2}
  \fmfright{r1}
  \fmf{boson}{l1,v1} 
  \fmf{boson}{l2,v1}
  \fmf{boson}{v1,r1}
 \end{fmfgraph}
\end{fmffile}
\\ \noindent
and one describes the interaction of four photons, which reads 

\begin{eqnarray}
v^{\mu\nu\lambda\rho}=\beta\left(-2 k_{2\sigma}k_4^{\sigma}\delta^{\mu\nu}\delta^{\lambda\rho}
+2 k_{2}^{\rho}k_4^{\lambda} \delta^{\mu\nu} 
+2 k_{2}^{\mu}k_4^{\nu} \delta^{\lambda\rho}
\right.\nonumber\\\left.
-2 k_{2}^{\mu}k_4^{\lambda}\delta^{\nu\rho}
-2 k_{3\sigma}k_4^{\sigma}\delta^{\mu\nu}\delta^{\lambda\rho}
+2 k_{3}^{\rho}k_4^{\lambda}\delta^{\mu\nu}\right).
\end{eqnarray}

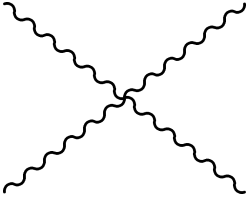
\begin{fmffile}{gQED7}
 \begin{fmfgraph}(40,25)
  \fmfleft{l1,l2}
  \fmfright{r1,r2}
  \fmf{boson}{l1,v1} 
  \fmf{boson}{l2,v1}
  \fmf{boson}{v1,r1}
  \fmf{boson}{v1,r2}  
 \end{fmfgraph}
\end{fmffile}

\section{Generalized Gauge Principle in General Relativity}

Since general relativity can also be formulated as a gauge theory, the generalization of the gauge principle should also 
have an influence on the description of gravity.
It is important that in such a consideration, where gravity is modified according to a generalized uncertainty principle,
the emergence of a minimal length in the description of quantum field theories cannot be interpreted as an approximative 
treatment of properties arising from a quantum description of gravity on a fundamental level. That means that the minimal
length is not induced by an underlying quantum theory of gravity, but the causal connection is the other way round. The
description of gravity, even on a classical level, is influenced by the assumption of a generalized uncertainty principle.
Therefore the generalized uncertainty principle should be interpreted as a fundamental description of nature in this context or
one has at least to assume that it has another origin. In \cite{Calmet:2010tx} and \cite{Percacci:2010af} the possibility is
argued that a minimal length could indeed arise from completely different scenarios being independent of the
usual considerations regarded as an effective description of the consequences of a quantum theory of gravity.      
Concerning the gauge theoretic description of general relativity there can be used the translation group as well as the Lorentz
group $SO(3,1)$ as gauge group. The first possibility leads to a formulation of general relativity where the torsion as a
decisive quantity plays the central role \cite{Cho:1975dh},\cite{Hehl:1976kj},\cite{Maluf:2003fs} but which is equivalent 
to usual general relativity.
Here it will be considered the $SO(3,1)$ gauge theory \cite{Ramond:1990},\cite{Hehl:1976kj} and it will be generalized 
to the case of matter field equations obeying a generalized uncertainty principle. This means that there is 
postulated invariance of the matter action under local Lorentz transformations

\begin{eqnarray}
x^m  \rightarrow  \Lambda^m_n(x) x^n,\quad\quad 
\psi \rightarrow  U(x)\psi=\exp\left(\frac{i}{2}\Lambda^{ab}(x)\Sigma_{ab}\right)\psi,
\label{local_Lorentz_transformation}
\end{eqnarray}
where the $\Sigma_{ab}$ describe the generators of the Lorentz group represented within the Dirac spinor space 
$\Sigma_{ab}=-\frac{i}{4}\left[\gamma_a,\gamma_b\right]$, fulfilling the commutation relation
$\left[\Sigma_{ab},\Sigma_{cd}\right]=\eta_{bc}\Sigma_{ad}-\eta_{ac}\Sigma_{bd}+\eta_{bd}\Sigma_{ca}-\eta_{ad}\Sigma_{cb}$,
where $\eta_{mn}$ describes the flat Minkowski metric.
Within the usual formulation one has to introduce the following covariant derivative:

\begin{equation}
D_m=e^\mu_m\left(\partial_\mu{\bf 1}
+\frac{i}{2}\omega_\mu^{ab}\Sigma_{ab}\right)=e^\mu_m D_\mu,
\label{usual_covariant_derivative_gravity}
\end{equation} 
where the spin connection $\omega_\mu^{ab}$ is related to the tetrad $e^\mu_m$ according to

\begin{equation}
\omega_\mu^{ab}=2e^{\nu a}\partial_\mu e_\nu^b-2e^{\nu b} \partial_\mu e_\nu^a
-2e^{\nu a}\partial_\nu e_\mu^b+2e^{\nu b} \partial_\nu e_\mu^a
+e_{\mu c} e^{\nu a}e^{\sigma b}\partial_\sigma e_\nu^c-e_{\mu c} e^{\nu a}e^{\sigma b}\partial_\nu e_\sigma^c
\label{connection_tetrad}
\end{equation}
and the tetrad $e^m_\mu$ is related to the general metric $g_{\mu\nu}$  as usual $g_{\mu\nu}=e_\mu^m e_\nu^n \eta_{mn}$. Note
that latin indices denote as usual flat Minkowski coordinates, whereas greek indices denote curved coordinates. The covariant
derivative ($\ref{usual_covariant_derivative_gravity}$) transforms according to

\begin{equation}
D_m \rightarrow \Lambda_m^n(x) U(x) D_n U^{\dagger}(x),
\label{transformation_covariant_derivative_gravity}
\end{equation}
since $e_m^\mu \rightarrow \Lambda_m^n(x) e_n^\mu$ and $\omega_\mu \rightarrow U^{\dagger}(x)\omega_\mu
U(x)-U^{\dagger}(x)\partial_\mu U(x)$. 
This means that in the generalized case one obtains the following matter action being invariant under a local $SO(3,1)$ 
gauge transformation consisting of ($\ref{local_Lorentz_transformation}$) and the corresponding transformation
($\ref{transformation_covariant_derivative_gravity}$):

\begin{eqnarray}
\mathcal{S}_{m}=\int d^4 x \sqrt{-g} \bar \psi \left[i\gamma^m e_m^\mu\left({\bf 1}-\beta D^{\rho} D_{\rho}\right)D_\mu-m \right]\psi
=\int d^4 x\ e \bar \psi \left[i\gamma^m e_m^\mu \mathcal{D}_\mu-m \right]\psi,
\label{generalized_matter_action_gravity}
\end{eqnarray}
where $g=\det\left[g_{\mu\nu}\right]$, $e=\det\left[e_\mu^m\right]$, and there is introduced the generalized covariant 
derivative according to ($\ref{definition_generalized_covariant_derivative}$):

\begin{equation}
\mathcal{D}_m=e_m^\mu\left({\bf 1}-\beta D^{\rho} D_{\rho}\right)D_\mu.
\label{generalized_covariant_derivative_gravity}
\end{equation}
Because of the generalized coupling of the gauge field to the matter field, ($\ref{generalized_matter_action_gravity}$)
contains a generalized equivalence principle. If the generalized field strength tensor corresponding 
to the generalized covariant derivative ($\ref{generalized_covariant_derivative_gravity}$) is built, one obtains

\begin{eqnarray}
\mathcal{F}_{mn}&=&\left[\mathcal{D}_m,\mathcal{D}_n\right]=
\left[e_m^\mu \mathcal{D}_\mu,e_n^\nu\mathcal{D}_\nu\right]
=e_m^\mu\left(\mathcal{D}_\mu e_n^\nu\right)\mathcal{D}_\nu
-e_n^\mu\left(\mathcal{D}_\mu e_m^\nu\right)\mathcal{D}_\nu
+e_m^\mu e_n^\nu \mathcal{D}_\mu \mathcal{D}_\nu-e_n^\mu e_m^\nu \mathcal{D}_\mu \mathcal{D}_\nu\nonumber\\
&=&e_m^\mu e_n^\nu \mathcal{T}_{\mu\nu}^p D_p+e_m^\mu e_n^\nu \frac{i}{2}\mathcal{R}_{\mu\nu}^{ab}\Sigma_{ab}
=\mathcal{T}_{mn}^p D_p+\frac{i}{2}\mathcal{R}_{mn}^{ab}\Sigma_{ab},
\end{eqnarray}
where $\mathcal{T}_{\mu\nu}^p$ and $\mathcal{R}_{\mu\nu}^{ab}$ are the generalized quantities corresponding to the usual 
torsion tensor and the usual Riemann curvature tensor. If the torsion is assumed to vanish $\mathcal{T}_{mn}^p=0$, 
there remains the generalized Riemann tensor $\mathcal{R}_{\mu\nu}^{ab}$ 
reading in analogy to ($\ref{generalized_tensor}$) as follows:

\begin{equation}
\mathcal{R}_{\mu\nu}^{ab}=R_{\mu\nu}^{ab}-2\beta D^\rho D_\rho R_{\mu\nu}^{ab}
-\beta\left(D^\rho R_{\mu\rho}^{ab}D_\nu-D^\rho R_{\nu\rho}^{ab} D_\mu\right)
-\beta\left(R_{\mu\rho}^{ab}D^\rho D_\nu-R_{\nu\rho}^{ab}D^\rho D_\mu\right)+\mathcal{O}\left(\beta^2\right),
\label{generalized_Riemann}
\end{equation}
where $R_{\mu\nu}^{ab}$ is the usual Riemann tensor expressed with two Minkowski indices reading

\begin{equation}
R_{\mu\nu}^{ab}=\partial_\mu \omega_\nu^{ab}-\partial_\nu \omega_\mu^{ab}
+\omega_\mu^{ac}\omega_\nu^{cb}-\omega_\nu^{ac}\omega_\mu^{cb}.
\end{equation}

\subsection{Generalized Einstein Hilbert Action and Corresponding Field Equation}

From ($\ref{generalized_Riemann}$) the generalization of the Einstein Hilbert action has to be built. Therefore
the generalized Ricci scalar has to be built according to

\begin{equation}
\mathcal{R}=e^\mu_a e^\nu_b \mathcal{R}_{\mu\nu}^{ab}
=e^\mu_a e^\nu_b\left[R_{\mu\nu}^{ab}-2\beta D^\rho D_\rho R_{\mu\nu}^{ab}
-\beta\left(D^\rho R_{\mu\rho}^{ab}D_\nu-D^\rho R_{\nu\rho}^{ab} D_\mu\right)
-\beta\left(R_{\mu\rho}^{ab}D^\rho D_\nu-R_{\nu\rho}^{ab}D^\rho D_\mu\right)\right].
\end{equation}
Since $R_{\mu\nu}^{ab}=-R_{\mu\nu}^{ba}$ and $D_\mu e_\nu^a=0$, if the torsion vanishes, this expression can 
be transformed to

\begin{equation} 
\mathcal{R}=R-2\beta D^\rho D_\rho R-2\beta D^\rho e^\nu_b R_{\rho}^{b} D_\nu
-2\beta e^\nu_b R_{\rho}^{b}D^\rho D_\nu.
\end{equation}
This leads to the following generalized Einstein Hilbert action:

\begin{equation} 
\mathcal{S}_{EH}=\frac{1}{2\kappa}\int d^4 x \sqrt{-g} \mathcal{R}
=\frac{1}{2\kappa}\int d^4 x\ e
\left[R-2\beta D^\rho D_\rho R
-2\beta D^\rho e^\nu_b R_{\rho}^{b} D_\nu
-2\beta e^\nu_b R_{\rho}^{b}D^\rho D_\nu \right]
\label{generalized_Einstein_Hilbert}
\end{equation}
where $\kappa=8\pi G$ and $G$ denotes the gravitational constant.
To obtain the corresponding free field equations for the gravitational field, the variation of the generalized Einstein Hilbert action ($\ref{generalized_Einstein_Hilbert}$) has to be built with respect to the tetrad $e^\mu_a$. 
If $R_\beta$ is defined according to $\mathcal{R}\equiv R+R_\beta+\mathcal{O}\left(\beta^2\right)$, then the 
variation can be expressed as 

\begin{equation}
\delta \mathcal{S}_{EH}=\frac{1}{2\kappa}\int d^4 x\ \delta \left[e \mathcal{R}\right]
=\frac{1}{2\kappa}\int d^4 x\ \delta \left[e R+e R_\beta\right]
=\frac{1}{2\kappa}\int d^4 x \left[\delta \left(e R\right)+\delta \left(e R_\beta\right)\right].
\end{equation}
The first term yields, of course, the usual Einstein equation, whereas the second term yields the corrections according
to the generalized uncertainty principle. To determine the contribution of the correction term, there has to be calculated 
the variation $\delta \left[e R_\beta\right]$ 

\begin{eqnarray}
\frac{1}{2\kappa}\int d^4 x\ \delta\left[e R_\beta\right]&=&
-\frac{\beta}{\kappa}\int d^4 x \delta \left[e D^\rho D_\rho e^\mu_a e^\nu_b R_{\mu\nu}^{ab}
+e D^\rho e^\mu_a e^\nu_b R_{\mu\rho}^{ab} D_\nu +e e^\mu_a e^\nu_bR_{\mu\rho}^{ab} D^\rho D_\nu \right],\nonumber\\
&=&-\frac{\beta}{\kappa}\int d^4 x\ \delta \left[D^\rho D_\rho e e^\mu_a e^\nu_b R_{\mu\nu}^{ab}
+D^\rho e e^\mu_a e^\nu_b R_{\mu\rho}^{ab} D_\nu +e e^\mu_a e^\nu_bR_{\mu\rho}^{ab} D^\rho D_\nu \right],\nonumber\\
&=&-\frac{\beta}{\kappa}\int d^4 x\ \partial^\rho \delta \left[D_\rho e e^\mu_a e^\nu_b R_{\mu\nu}^{ab}
+e e^\mu_a e^\nu_b R_{\mu\rho}^{ab} D_\nu\right]-\frac{\beta}{\kappa} \int d^4 x\ \delta 
\left[e e^\mu_a e^\nu_bR_{\mu\rho}^{ab} D^\rho D_\nu \right],\nonumber\\
&=&\underbrace{-\frac{\beta}{\kappa}\oint\limits_{\partial V} d^3 x\ \delta \left[D_\rho e e^\mu_a e^\nu_b R_{\mu\nu}^{ab}
+e e^\mu_a e^\nu_b R_{\mu\rho}^{ab} D_\nu\right]}_{=0}-\frac{\beta}{\kappa} \int d^4 x\ \delta 
\left[e e^\mu_a e^\nu_bR_{\mu\rho}^{ab} D^\rho D_\nu \right],\nonumber\\
&=&-\frac{\beta}{\kappa}\int d^4 x\ \delta \left[e e^\mu_a e^\nu_bR_{\mu\rho}^{ab} D^\rho D_\nu \right],
\end{eqnarray}
where Stokes theorem $\int_{V} dA=\int_{\partial V} A$ has been used in the forth step and the fact that the variation 
vanishes at infinity in the fifth step makes the surface term vanishing $-\frac{\beta}{\kappa}\oint_{\partial V} d^3 x\ \delta
\left[D_\rho e e^\mu_a e^\nu_b R_{\mu\nu}^{ab}+e e^\mu_a e^\nu_b R_{\mu\rho}^{ab} D_\nu\right]=0$. 
Variation of the remaining term yields

\begin{eqnarray}
-\frac{\beta}{\kappa}\int d^4 x\ \delta\left[e e^\mu_a e^\nu_bR_{\mu\rho}^{ab} D^\rho D_\nu \right]
&=&-\frac{\beta}{\kappa}\int d^4 x \left[-e e^a_\mu \delta e^\mu_a R D^\rho D_\nu
+e \delta e^\mu_a e^\nu_b R^{ab}_{\mu\rho} D^\rho D_\nu
+e \delta e^\mu_a R^a_\rho D^\rho D_\mu
+e e^\nu_b e^\lambda_c \delta\left(R_{\lambda\rho}^{cb} D^\rho D_\nu\right)\right],\nonumber\\
\label{variation_correction_term}
\end{eqnarray}
if it is used that $\delta e=-e e_\mu^a \delta e^\mu_a$.
This means that the variation of the last term has to be calculated with respect to the tetrad $e_\mu^a$. 
First, the expression that is varied $R_{\lambda\rho}^{cb} D^\rho D_\nu$ has to be expressed in terms of the
connection $\omega_\mu^{ab}$

\begin{eqnarray}
R_{\lambda\rho}^{bc}D^\rho D_\nu&=&\left(\partial_\lambda \omega_\rho^{bc} \partial^{\rho} \omega_{\nu}^{ef}
-\partial_\rho \omega_\lambda^{bc} \partial^{\rho} \omega_{\nu}^{ef}
+\partial_\lambda \omega_{\rho}^{bc} \omega^{\rho eg}\omega_\nu^{gf}
-\partial_\rho \omega_{\lambda}^{bc} \omega^{\rho eg}\omega_\nu^{gf}\nonumber\right.\\
&&\left.+\omega_{\lambda}^{bd} \omega_\rho^{dc}\partial^\rho \omega_\nu^{ef}
-\omega_{\rho}^{bd} \omega_\lambda^{dc}\partial^\rho \omega_\nu^{ef}
+\omega_\lambda^{bd}\omega_\rho^{dc} \omega^{\rho eg}\omega_{\nu}^{gf}
-\omega_\rho^{bd}\omega_\lambda^{dc} \omega^{\rho eg}\omega_{\nu}^{gf}\right)\Sigma_{ef}.
\label{expression_connection} 
\end{eqnarray}
If ($\ref{expression_connection}$) is inserted into ($\ref{variation_correction_term}$), the variation of the last term
with respect to the connection $\omega_\mu^{ab}$ yields

\begin{eqnarray}
&&-\frac{\beta}{\kappa}\int d^4 x \left[e e^\nu_c e^\lambda_b \delta \left(R_{\lambda\rho}^{bc}D^\rho D_\nu\right)\right]
\nonumber\\
&=&-\frac{\beta}{\kappa}\int d^4 x \left[e e^\nu_c e^\lambda_b 
\left(\partial_\lambda \delta \omega_\rho^{bc} \partial^{\rho} \omega_{\nu}^{ef}
+\partial_\lambda \omega_\rho^{bc} \partial^{\rho} \delta \omega_{\nu}^{ef}
-\partial_\rho \delta \omega_\lambda^{bc} \partial^{\rho} \omega_{\nu}^{ef}
-\partial_\rho \omega_\lambda^{bc} \partial^{\rho} \delta \omega_{\nu}^{ef}
+\partial_\lambda \delta \omega_{\rho}^{bc} \omega^{\rho eg}\omega_\nu^{gf}
\right.\right.
\nonumber\\
&&\left.\left.
+\partial_\lambda \omega_{\rho}^{bc} \delta \omega^{\rho eg}\omega_\nu^{gf}
+\partial_\lambda \omega_{\rho}^{bc} \omega^{\rho eg}\delta \omega_\nu^{gf}
-\partial_\rho \delta \omega_{\lambda}^{bc} \omega^{\rho eg}\omega_\nu^{gf}
-\partial_\rho \omega_{\lambda}^{bc} \delta \omega^{\rho eg}\omega_\nu^{gf}
-\partial_\rho \omega_{\lambda}^{bc} \omega^{\rho eg}\delta \omega_\nu^{gf}
\right.\right.
\nonumber\\
&&\left.\left.
+\delta \omega_{\lambda}^{bd} \omega_\rho^{dc}\partial^\rho \omega_\nu^{ef}
+\omega_{\lambda}^{bd} \delta \omega_\rho^{dc}\partial^\rho \omega_\nu^{ef}
+\omega_{\lambda}^{bd} \omega_\rho^{dc}\partial^\rho \delta \omega_\nu^{ef}
-\delta \omega_{\rho}^{bd} \omega_\lambda^{dc}\partial^\rho \omega_\nu^{ef}
-\omega_{\rho}^{bd} \delta \omega_\lambda^{dc}\partial^\rho \omega_\nu^{ef}
\right.\right.
\nonumber\\
&&\left.\left.
-\omega_{\rho}^{bd} \omega_\lambda^{dc}\partial^\rho \delta \omega_\nu^{ef}
+\delta \omega_\lambda^{bd}\omega_\rho^{dc} \omega^{\rho eg}\omega_{\nu}^{gf}
+\omega_\lambda^{bd}\delta \omega_\rho^{dc} \omega^{\rho eg}\omega_{\nu}^{gf}
+\omega_\lambda^{bd}\omega_\rho^{dc} \delta \omega^{\rho eg}\omega_{\nu}^{gf}
+\omega_\lambda^{bd}\omega_\rho^{dc} \omega^{\rho eg}\delta \omega_{\nu}^{gf}
\right.\right.
\nonumber\\
&&\left.\left.
-\delta \omega_\rho^{bd}\omega_\lambda^{dc} \omega^{\rho eg}\omega_{\nu}^{gf}
-\omega_\rho^{bd}\delta \omega_\lambda^{dc} \omega^{\rho eg}\omega_{\nu}^{gf}
-\omega_\rho^{bd}\omega_\lambda^{dc} \delta \omega^{\rho eg}\omega_{\nu}^{gf}
-\omega_\rho^{bd}\omega_\lambda^{dc} \omega^{\rho eg}\delta \omega_{\nu}^{gf}\right)\Sigma_{ef}\right]\nonumber\\
&\equiv&-\frac{\beta}{\kappa}\int d^4 x \left[\Xi_{hi}^{\kappa} \delta \omega^{hi}_\kappa \right],
\label{variation_with_respect_connection}
\end{eqnarray}
where the expression $\Xi_{hi}^{\kappa}$ has been defined according to

\begin{eqnarray}
\Xi_{hi}^{\kappa}&=&
\left[-\delta_{\rho}^{\kappa} \delta_{h}^{b} \delta_{i}^{c} \partial_\lambda\left(e e^\nu_c e^\lambda_b 
\partial^{\rho} \omega_{\nu}^{ef}\right)
-\delta_{\nu}^{\kappa} \delta_{h}^{e} \delta_{i}^{f} \partial^{\rho} \left(e e^\nu_c e^\lambda_b \partial_\lambda \omega_\rho^{bc}\right) 
+\delta_{\lambda}^{\kappa} \delta_{h}^{b} \delta_{i}^{c} \partial_\rho \left(e e^\nu_c e^\lambda_b 
\partial^{\rho} \omega_{\nu}^{ef}\right)
+\delta_{\nu}^{\kappa} \delta_{h}^{e} \delta_{i}^{f} \partial^{\rho} \left(e e^\nu_c e^\lambda_b \partial_\rho \omega_\lambda^{bc}\right)
\right.\label{definition_capital_xi}\\&&\left.
-\delta_{\rho}^{\kappa} \delta_{h}^{b} \delta_{i}^{c} \partial_\lambda \left(e e^\nu_c e^\lambda_b 
\omega^{\rho eg}\omega_\nu^{gf}\right)
+\delta^{\rho\kappa} \delta_{h}^{e} \delta_{i}^{g} e e^\nu_c e^\lambda_b \partial_\lambda \omega_{\rho}^{bc} 
\omega_\nu^{gf}
+\delta_{\nu}^{\kappa} \delta_{h}^{g} \delta_{i}^{f} e e^\nu_c e^\lambda_b \partial_\lambda \omega_{\rho}^{bc} \omega^{\rho eg}
+\delta_{\lambda}^{\kappa} \delta_{h}^{b} \delta_{i}^{c} \partial_\rho \left(e e^\nu_c e^\lambda_b 
\omega^{\rho eg}\omega_\nu^{gf}\right)
\right.\nonumber\\&&\left.
-\delta^{\rho\kappa} \delta_{h}^{e} \delta_{i}^{g} e e^\nu_c e^\lambda_b \partial_\rho \omega_{\lambda}^{bc} 
\omega_\nu^{gf}
-\delta_{\nu}^{\kappa} \delta_{h}^{g} \delta_{i}^{f} e e^\nu_c e^\lambda_b \partial_\rho \omega_{\lambda}^{bc} \omega^{\rho eg}
+\delta_{\lambda}^{\kappa} \delta_{h}^{b} \delta_{i}^{d} e e^\nu_c e^\lambda_b 
\omega_\rho^{dc}\partial^\rho \omega_\nu^{ef}
+\delta_{\rho}^{\kappa} \delta_{h}^{d} \delta_{i}^{c} e e^\nu_c e^\lambda_b \omega_{\lambda}^{bd} 
\partial^\rho \omega_\nu^{ef}
\right.\nonumber\\&&\left.
-\delta_{\nu}^{\kappa} \delta_{h}^{e} \delta_{i}^{f}\partial^\rho
\left(e e^\nu_c e^\lambda_b \omega_{\lambda}^{bd} \omega_\rho^{dc}\right)
-\delta_{\rho}^{\kappa} \delta_{h}^{b} \delta_{i}^{d} e e^\nu_c e^\lambda_b
\omega_\lambda^{dc}\partial^\rho \omega_\nu^{ef}
-\delta_{\lambda}^{\kappa} \delta_{h}^{d} \delta_{i}^{c} e e^\nu_c e^\lambda_b \omega_{\rho}^{bd}
\partial^\rho \omega_\nu^{ef}
+\delta_{\nu}^{\kappa} \delta_{h}^{e} \delta_{i}^{f} \partial^\rho \left(e e^\nu_c e^\lambda_b \omega_{\rho}^{bd} \omega_\lambda^{dc}\right)
\right.\nonumber\\&&\left.
+\delta_{\lambda}^{\kappa} \delta_{h}^{b} \delta_{i}^{d} e e^\nu_c e^\lambda_b 
\omega_\rho^{dc} \omega^{\rho eg}\omega_{\nu}^{gf}
+\delta_{\rho}^{\kappa} \delta_{h}^{d} \delta_{i}^{c} e e^\nu_c e^\lambda_b \omega_\lambda^{bd}
\omega^{\rho eg}\omega_{\nu}^{gf}
+\delta^{\rho\kappa} \delta_{h}^{e} \delta_{i}^{g} e e^\nu_c e^\lambda_b \omega_\lambda^{bd}\omega_\rho^{dc} 
\omega_{\nu}^{gf}
+\delta_{\nu}^{\kappa} \delta_{h}^{g}\delta_{i}^{f} e e^\nu_c e^\lambda_b \omega_\lambda^{bd}\omega_\rho^{dc} \omega^{\rho eg}
\right.\nonumber\\&&\left.
-\delta_{\rho}^{\kappa} \delta_{h}^{b} \delta_{i}^{d} e e^\nu_c e^\lambda_b 
\omega_\lambda^{dc} \omega^{\rho eg}\omega_{\nu}^{gf}
-\delta_{\lambda}^{\kappa} \delta_{h}^{d} \delta_{i}^{c} e e^\nu_c e^\lambda_b \omega_\rho^{bd}
\omega^{\rho eg}\omega_{\nu}^{gf}
-\delta^{\rho\kappa} \delta_{h}^{e} \delta_{i}^{g} e e^\nu_c e^\lambda_b \omega_\rho^{bd}\omega_\lambda^{dc} 
\omega_{\nu}^{gf}
-\delta_{\nu}^{\kappa} \delta_{h}^{g} \delta_{i}^{f} e e^\nu_c e^\lambda_b \omega_\rho^{bd}\omega_\lambda^{dc} \omega^{\rho eg}
\right]\Sigma_{ef}\nonumber
\end{eqnarray}
and used again the fact that the variation vanishes at infinity.
To obtain the variation of ($\ref{variation_with_respect_connection}$) with respect to the tetrad $e_\mu^a$,
($\ref{connection_tetrad}$) has to be used. Variation of ($\ref{connection_tetrad}$) with respect to $e_\mu^a$ yields

\begin{eqnarray}
\delta \omega_\mu^{ab}&=&2 \delta e^{\nu a}\partial_\mu e_\nu^b
+2e^{\nu a}\partial_\mu \delta e_\nu^b
-2\delta e^{\nu b} \partial_\mu e_\nu^a
-2e^{\nu b} \partial_\mu \delta e_\nu^a
-2\delta e^{\nu a}\partial_\nu e_\mu^b
-2e^{\nu a}\partial_\nu \delta e_\mu^b
+2\delta e^{\nu b} \partial_\nu e_\mu^a
+2e^{\nu b} \partial_\nu \delta e_\mu^a\nonumber\\
&&+\delta e_{\mu c} e^{\nu a}e^{\sigma b}\partial_\sigma e_\nu^c
+e_{\mu c} \delta e^{\nu a}e^{\sigma b}\partial_\sigma e_\nu^c
+e_{\mu c} e^{\nu a}\delta e^{\sigma b}\partial_\sigma e_\nu^c
+e_{\mu c} e^{\nu a}e^{\sigma b}\partial_\sigma \delta e_\nu^c\nonumber\\
&&-\delta e_{\mu c} e^{\nu a}e^{\sigma b}\partial_\nu e_\sigma^c
-e_{\mu c} \delta e^{\nu a}e^{\sigma b}\partial_\nu e_\sigma^c
-e_{\mu c} e^{\nu a}\delta e^{\sigma b}\partial_\nu e_\sigma^c
-e_{\mu c} e^{\nu a}e^{\sigma b}\partial_\nu \delta e_\sigma^c.
\label{variation_connection}
\end{eqnarray}
If ($\ref{variation_connection}$) is inserted into ($\ref{variation_with_respect_connection}$), one obtains the variation with
respect to the tetrad

\begin{eqnarray}
-\frac{\beta}{\kappa}\int d^4 x \left[\Xi_{hi}^{\kappa} \delta \omega^{hi}_\kappa \right]\equiv-\frac{\beta}{\kappa}\int d^4 x \left[\Theta_\mu^a \delta e^\mu_a \right],
\end{eqnarray}
where $\Theta_\mu^a$ is defined according to

\begin{eqnarray}
\Theta_\mu^a&=&2\delta_\mu^\nu \delta^{ah} \Xi_{hi}^{\kappa}
\partial_\kappa e_\nu^i
-2\delta_{\mu\nu}\delta^{ai}\partial_\kappa \left(\Xi_{hi}^{\kappa}e^{\nu h}\right)
-2\delta_\mu^{\nu}\delta^{ai} \Xi_{hi}^{\kappa}
\partial_\kappa e_\nu^h
+2\delta_{\mu\nu}\delta^{ah} \partial_\kappa \left(\Xi_{hi}^{\kappa}e^{\nu i}\right) 
\nonumber\\
&&-2\delta_\mu^\nu \delta^{ah} \Xi_{hi}^{\kappa}
\partial_\nu e_\kappa^i
+2\delta_{\mu\kappa} \delta^{ai} \partial_\nu \left(\Xi_{hi}^{\kappa}e^{\nu h}\right)
+2\delta_\mu^\nu \delta^{ai} \Xi_{hi}^{\kappa}
\partial_\nu e_\kappa^h
-2\delta_{\mu\kappa}\delta^{ah} \partial_\nu\left(\Xi_{hi}^{\kappa}e^{\nu i}\right)
\nonumber\\
&&+\delta_{\mu\kappa}\delta_c^{a} \Xi_{hi}^{\kappa}
e^{\nu h}e^{\sigma i}\partial_\sigma e_\nu^c
+\delta_\mu^\nu \delta^{ah} \Xi_{hi}^{\kappa}e_{\kappa c} 
e^{\sigma i}\partial_\sigma e_\nu^c
+\delta_\mu^\sigma \delta^{ai} \Xi_{hi}^{\kappa}e_{\kappa c} e^{\nu h}
\partial_\sigma e_\nu^c
-\delta_{\mu\nu}\delta^{ac}\partial_\sigma\left(\Xi_{hi}^{\kappa}e_{\kappa c} e^{\nu h}e^{\sigma i}\right)
\nonumber\\
&&-\delta_{\mu\kappa} \delta_c^a\Xi_{hi}^{\kappa}
e^{\nu h}e^{\sigma i}\partial_\nu e_\sigma^c
-\delta_\mu^\nu \delta^{ah} \Xi_{hi}^{\kappa}e_{\kappa c}
e^{\sigma i}\partial_\nu e_\sigma^c
-\delta_\mu^\sigma \delta^{ai} \Xi_{hi}^{\kappa}e_{\kappa c} e^{\nu h}
\partial_\nu e_\sigma^c
+\delta_{\mu\sigma} \delta^{ac} \partial_\nu \left(\Xi_{hi}^{\kappa}e_{\kappa c} e^{\nu h}e^{\sigma i}\right).
\label{definition_capital_theta}
\end{eqnarray}
This means that the variation ($\ref{variation_correction_term}$) now reads

\begin{eqnarray}
-\frac{\beta}{\kappa}\int d^4 x \delta\left[e e^\mu_a e^\nu_bR_{\mu\rho}^{ab} D^\rho D_\nu \right]
&=&-\frac{\beta}{\kappa}\int d^4 x \left[-e e^a_\mu \delta e^\mu_a R D^\rho D_\nu
+e \delta e^\mu_a e^\nu_b R^{ab}_{\mu\rho} D^\rho D_\nu
+e \delta e^\mu_a R^a_\rho D^\rho D_\mu
+\Theta_\mu^a \delta e^\mu_a \right],
\end{eqnarray}
which means that variation of the complete action yields the following generalized vacuum Einstein field equation:

\begin{equation}
R_\mu^a-\frac{1}{2} R e_\mu^a-\beta\left[-e^a_\mu R D^\rho D_\nu+e^\nu_b R^{ab}_{\mu\rho} D^\rho D_\nu
+R^a_\rho D^\rho D_\mu+\frac{1}{e}\Theta^a_\mu\right]=0,
\end{equation}
where $\Theta_\mu^a$ is defined according to ($\ref{definition_capital_theta}$) and ($\ref{definition_capital_xi}$).
If the matter action ($\ref{generalized_matter_action_gravity}$) of the fermionic field is included and the complete
gravity action

\begin{equation}
\mathcal{S}_{gr}=\mathcal{S}_{m}+\mathcal{S}_{EH}=\int d^4 x\ e \bar \psi \left[i\gamma^m e_m^\mu \mathcal{D}_\mu
-m \right]\psi
+\frac{1}{2\kappa}\int d^4 x\ e \mathcal{R}
\end{equation}
is varied with respect to the tetrad $\frac{\delta \mathcal{S}_{gr}}{\delta e^\mu_a}$, one obtains

\begin{eqnarray}
R_\mu^a-\frac{1}{2} R e_\mu^a-\beta\left[-e^a_\mu R D^\rho D_\nu+e^\nu_b R^{ab}_{\mu\rho} D^\rho D_\nu
+R^a_\rho D^\rho D_\mu+
\frac{1}{e}\Theta^a_\mu
\right]=-\kappa T_\mu^a,
\end{eqnarray}
where the energy momentum tensor is defined as usual as $T_\mu^a=\frac{1}{e}\frac{\delta \mathcal{S}_m}{\delta e^\mu_a}$. 

\subsection{Generalized Energy Momentum Tensor for a Dirac Field Coupled to the Gravitational Field}

To obtain the concrete expression of the energy momentum tensor in case of a fermionic field that is coupled to the
gravitational field according to the generalized gauge theory presented here, the expression
($\ref{generalized_matter_action_gravity}$) has to be varied with respect to the tetrad $e^\mu_a$. Accordingly, the 
variation of the matter action with respect to the tetrad $e^\mu_a$ reads

\begin{eqnarray}
\delta \mathcal{S}_{m}
&=&\delta \int d^4 x\ e \bar \psi \left[i\gamma^m e_m^\mu \mathcal{D}_\mu-m \right]\psi
=\delta \int d^4 x e\ \bar \psi \left[i\gamma^m e_m^\mu \left({\bf 1}-\beta D^{\rho} D_{\rho}\right)D_\mu-m \right]\psi\nonumber\\
&=&\int d^4 x \left\{-e e^m_\mu \delta e_m^\mu \bar \psi \left[i\gamma^m e_m^\mu \left({\bf 1}-\beta D^{\rho} D_{\rho}\right)D_\mu-m \right]\psi
+e \bar \psi i\gamma^m \delta e_m^\mu \left({\bf 1}-\beta D^{\rho} D_{\rho}\right)D_\mu\psi\right.\nonumber\\
&&\left.+e \bar \psi i\gamma^m e_m^\mu \delta \left[\left({\bf 1}-\beta D^{\rho} D_{\rho}\right)D_\mu\right]\psi\right\}.
\label{variation_matter}
\end{eqnarray}
To calculate the last term, it has to be expressed by the connection $\omega_\mu^{ab}$ and the connection has to be expressed
by the tetrad $e^\mu_a$

\begin{eqnarray}
&&\int d^4 x\ e \bar \psi i\gamma^m e_m^\mu\delta \left[\left({\bf 1}-\beta D^{\rho} D_{\rho}\right)D_\mu\right]\psi\nonumber\\
&=&\int d^4 x\ e \bar \psi i\gamma^m e_m^\mu \delta \left[i\omega_\mu^{ab}
-\beta\left(
i\partial^\rho \partial_\rho \omega_\mu^{ab}
+2i\partial^\rho \omega_\mu^{ab} \partial_\rho 
+i\omega_\mu^{ab} \partial_\rho \partial^\rho  
+i\partial^{\rho}\omega_\rho^{ab}\partial_\mu
+2i\omega^{\rho ab}\partial_\rho \partial_\mu
-\partial^\rho \omega_\rho^{ac}\omega_\mu^{cb}
\right.\right.\nonumber\\&&\left.\left.
-2\omega_\rho^{ac}\partial^\rho \omega_\mu^{cb}
-2\omega_\rho^{ac}\omega_\mu^{cb} \partial^\rho 
-\omega^{\rho ac}\omega_\rho^{cb}\partial_\mu
-i\omega^{\rho ac}\omega_\rho^{cd}\omega_\mu^{db}
\right)\right]\Sigma_{ab}\psi
\nonumber\\
&=&\int d^4 x\ e \bar \psi i\gamma^m e_m^\mu \left[i\delta \omega_\mu^{ab}
-\beta\left(
i\partial^\rho \partial_\rho \delta \omega_\mu^{ab}
+2i\partial^\rho \delta \omega_\mu^{ab} \partial_\rho 
+i\delta \omega_\mu^{ab} \partial^\rho \partial_\rho 
+i\partial^{\rho}\delta \omega_\rho^{ab}\partial_\mu
+2i\delta \omega^{\rho ab}\partial_\rho \partial_\mu
\right.\right.\nonumber\\&&\left.\left.
-\partial^\rho \delta \omega_\rho^{ac}\omega_\mu^{cb}
-\partial^\rho \omega_\rho^{ac}\delta \omega_\mu^{cb}
-2\delta \omega_\rho^{ac} \partial^\rho  \omega_\mu^{cb}
-2\omega_\rho^{ac}\partial^\rho \delta \omega_\mu^{cb}
-2\delta \omega_\rho^{ac}\omega_\mu^{cb} \partial^\rho 
-2\omega_\rho^{ac}\delta \omega_\mu^{cb}\partial^\rho 
-\delta \omega^{\rho ac}\omega_\rho^{cb}\partial_\mu
\right.\right.\nonumber\\&&\left.\left.
-\omega^{\rho ac}\delta \omega_\rho^{cb}\partial_\mu
-i\delta \omega^{\rho ac}\omega_\rho^{cd}\omega_\mu^{db}
-i\omega^{\rho ac}\delta \omega_\rho^{cd}\omega_\mu^{db}
-i\omega^{\rho ac}\omega_\rho^{cd}\delta \omega_\mu^{db}
\right)\right]\Sigma_{ab}\psi\nonumber\\
&\equiv&\int d^4 x \left[\xi_{hi}^{\kappa}\delta \omega_\kappa^{hi}\right]
\label{variation_with_respect_connection_2}
\end{eqnarray}
where $\xi_{hi}^{\kappa}$ has been defined according to

\begin{eqnarray}
\xi_{hi}^{\kappa}&=&\left\{
i\delta^{\kappa}_{\mu}\delta^{a}_{h}\delta^{b}_{i}A^\mu_{ab}
-\beta\left[
i\delta^{\kappa}_{\mu}\delta^{a}_{h}\delta^{b}_{i}\partial^\rho \partial_\rho A^\mu_{ab}
-2i\delta^{\kappa}_{\mu}\delta^{a}_{h}\delta^{b}_{i}\partial^\rho B^{\mu}_{\rho ab}
+i\delta^{\kappa}_{\mu}\delta^{a}_{h}\delta^{b}_{i}C^{\mu\rho}_{\rho ab}
-i\delta^{\kappa}_{\rho}\delta^{a}_{h}\delta^{b}_{i}\partial^{\rho}B^{\mu}_{\mu ab}
\right.\right.\nonumber\\&&\left.\left.
+2i\delta^{\kappa\rho}\delta^{a}_{h}\delta^{b}_{i}C^{\mu}_{\rho\mu ab}
+\delta^{\kappa}_{\rho}\delta^{a}_{h}\delta^{c}_{i}\partial^\rho 
\left(\omega_\mu^{cb}A^\mu_{ab}\right)
-\delta^{\kappa}_{\mu}\delta^{c}_{h}\delta^{b}_{i}\partial^\rho \omega_\rho^{ac}A^\mu_{ab}
-2\delta^{\kappa}_{\rho}\delta^{a}_{h}\delta^{c}_{i}
\partial^\rho \omega_\mu^{cb}A^\mu_{ab}
+2\delta^{\kappa}_{\mu}\delta^{c}_{h}\delta^{b}_{i}\partial^\rho \left(\omega_\rho^{ac}A^\mu_{ab}\right)
\right.\right.\nonumber\\&&\left.\left.
-2\delta^{\kappa}_{\rho}\delta^{a}_{h}\delta^{c}_{i}
\omega_\mu^{cb} B^{\mu\rho}_{ab} 
-2\delta^{\kappa}_{\mu}\delta^{c}_{h}\delta^{b}_{i}\omega_\rho^{ac}B^{\mu\rho}_{ab}
-\delta^{\kappa\rho}\delta^{a}_{h}\delta^{c}_{i}
\omega_\rho^{cb}B^{\mu}_{\mu ab}
-\delta^{\kappa}_{\rho}\delta^{c}_{h}\delta^{b}_{i}\omega^{\rho ac}B^{\mu}_{\mu ab}
-i\delta^{\kappa\rho}\delta^{a}_{h}\delta^{c}_{i}
\omega_\rho^{cd}\omega_\mu^{db}A^\mu_{ab}
\right.\right.\nonumber\\&&\left.\left.
-i\delta^{\kappa}_{\rho}\delta^{c}_{h}\delta^{d}_{i}\omega^{\rho ac}
\omega_\mu^{db}A^\mu_{ab}
-i\delta^{\kappa}_{\mu}\delta^{d}_{h}\delta^{b}_{i}\omega^{\rho ac}\omega_\rho^{cd}
A^\mu_{ab}\right]\right\}
\label{definition_lowercase_xi}
\end{eqnarray}
with $A^\mu_{ab}=e \bar \psi i\gamma^m e_m^\mu \Sigma_{ab}\psi$, 
$B^{\mu\nu}_{ab}=e \bar \psi i\gamma^m e_m^\mu \Sigma_{ab}\partial^\nu \psi$,
and $C^{\mu\nu}_{\rho ab}=e \bar \psi i\gamma^m e_m^\mu \Sigma_{ab}\partial^\nu \partial_\rho \psi$. 
If ($\ref{variation_connection}$) is inserted into ($\ref{variation_with_respect_connection_2}$), 
then one obtains

\begin{eqnarray}
\int d^4 x \left[\xi_{hi}^{\kappa} \delta \omega^{hi}_\kappa \right]
\equiv \int d^4 x \left[\theta_\mu^a \delta e^\mu_a \right]
\end{eqnarray}
with $\theta_\mu^a$ defined according to

\begin{eqnarray}
\theta_\mu^a&=&2\delta_\mu^\nu \delta^{ah} \xi_{hi}^{\kappa}
\partial_\kappa e_\nu^i
-2\delta_{\mu\nu}\delta^{ai}\partial_\kappa \left(\xi_{hi}^{\kappa}e^{\nu h}\right)
-2\delta_\mu^{\nu}\delta^{ai} \xi_{hi}^{\kappa}
\partial_\kappa e_\nu^h
+2\delta_{\mu\nu}\delta^{ah} \partial_\kappa \left(\xi_{hi}^{\kappa}e^{\nu i}\right)
\nonumber\\
&&-2\delta_\mu^\nu \delta^{ah} \xi_{hi}^{\kappa}
\partial_\nu e_\kappa^i
+2\delta_{\mu\kappa} \delta^{ai} \partial_\nu \left(\xi_{hi}^{\kappa}e^{\nu h}\right)
+2\delta_\mu^\nu \delta^{ai} \xi_{hi}^{\kappa}
\partial_\nu e_\kappa^h
-2\delta_{\mu\kappa}\delta^{ah} \partial_\nu\left(\xi_{hi}^{\kappa}e^{\nu i}\right)
\nonumber\\
&&+\delta_{\mu\kappa}\delta_c^{a} \xi_{hi}^{\kappa}
e^{\nu h}e^{\sigma i}\partial_\sigma e_\nu^c
+\delta_\mu^\nu \delta^{ah} \xi_{hi}^{\kappa}e_{\kappa c}
e^{\sigma i}\partial_\sigma e_\nu^c
+\delta_\mu^\sigma \delta^{ai} \xi_{hi}^{\kappa}e_{\kappa c} e^{\nu h}
\partial_\sigma e_\nu^c
-\delta_{\mu\nu}\delta^{ac}\partial_\sigma\left(\xi_{hi}^{\kappa}e_{\kappa c} e^{\nu h}e^{\sigma i}\right)
\nonumber\\
&&-\delta_{\mu\kappa} \delta_c^a\xi_{hi}^{\kappa}
e^{\nu h}e^{\sigma i}\partial_\nu e_\sigma^c
-\delta_\mu^\nu \delta^{ah} \xi_{hi}^{\kappa}e_{\kappa c} 
e^{\sigma i}\partial_\nu e_\sigma^c
-\delta_\mu^\sigma \delta^{ai} \xi_{hi}^{\kappa}e_{\kappa c} e^{\nu h}
\partial_\nu e_\sigma^c
+\delta_{\mu\sigma} \delta^{ac} \partial_\nu \left(\xi_{hi}^{\kappa}e_{\kappa c} 
e^{\nu h}e^{\sigma i}\right).
\label{definition_lowercase_theta}
\end{eqnarray} 
This means that the variation of the matter action ($\ref{variation_matter}$) reads now

\begin{eqnarray}
\delta \mathcal{S}_{m}
&=&\int d^4 x \left\{-e e^m_\mu \delta e_m^\mu \bar \psi \left[i\gamma^m e_m^\mu \left({\bf 1}-\beta D^{\rho} D_{\rho}\right)D_\mu-m \right]\psi
+e \bar \psi i\gamma^m \delta e_m^\mu \left({\bf 1}-\beta D^{\rho} D_{\rho}\right)D_\mu\psi,
+\theta_\mu^m \delta e^\mu_m \right\}
\end{eqnarray}
leading to the following energy momentum tensor:

\begin{equation}
T_\mu^a=\bar \psi i\gamma^a \left({\bf 1}-\beta D^{\rho} D_{\rho}\right)D_\mu 
\psi-e^a_\mu \bar \psi\left[i\gamma^m e_m^\nu \left({\bf 1}-\beta D^{\rho} D_{\rho}\right)D_\nu-m\right]\psi
+\theta_\mu^a,
\end{equation}
where $\theta_\mu^a$ is defined according ($\ref{definition_lowercase_theta}$) and ($\ref{definition_lowercase_xi}$).

\section{Summary and Discussion}

There has been considered an extension of the usual description of local gauge theories according to the implementation of 
a generalized uncertainty principle to quantum field theory, which causes generalized free field equations. Since a 
generalized uncertainty principle leads to additional derivative terms, if the momentum operator and the field equations 
are represented in position space, the description of a local gauge theory becomes more intricate. The reason lies in the 
fact that the additional derivative terms also have to be replaced by covariant derivatives to maintain local gauge invariance 
and therefore there appear additional terms containing the gauge potential. These terms describe a more complicated interaction
structure between the gauge field and the matter field. In accordance with this there can be defined a generalized covariant
derivative from which there is built a generalized field strength tensor. If the action of the gauge field is built by using 
this generalized definition of the field strength tensor, the dynamics of the gauge field contains additional self-interaction 
terms arising from this generalization. This means that even in case of an Abelian gauge theory like electrodynamics there
arises a self-interaction of the electromagnetic field. According to this, the generalized Feynman rules of quantum
electrodynamics have been determined by calculating the propagators under presupposition of the generalized momentum eigenstates
and extracting the vertices from the expressions of the action expressed explicitly in terms of the gauge potential. Because of
the additional interaction terms not only are the usual vertices changed but also completely new vertices arise. 
Since general relativity can also be formulated as a gauge theory, this generalization of gauge theories also changes the
structure of the gravitational interaction. The generalization of the covariant derivative in the $SO(3,1)$ gauge description
of general relativity causes a changing of the interaction of a matter field with the gravitational field, which can be seen as a modification of the equivalence principle. 
Further, as in the case of electrodynamics, the dynamics of the gravitational field itself is changed. The Einstein Hilbert
action is built from a generalized Riemann tensor defined as the commutator of the generalized covariant derivatives in
accordance with the field strength tensor in the case of electrodynamics, and the variation with respect to the tetrad field
yields the generalized Einstein equations. 
This means that the structure of the gravitational interaction and thus the dynamics of the metric structure of space-time
depend on the generalized uncertainty principle and thus on the corresponding smallest length within this consideration.
Usually the generalized uncertainty principle is seen as a consequence of a quantum description of gravity. In the present
paper the influence on gravity induced by a generalized uncertainty principle is considered instead, if it is assumed to yield
a fundamental description of nature or to have another origin. This means that a modified description of gravity is obtained
depending on the modification of the uncertainty principle independent of an underlying theory. However, the presented
modification of gravity could be seen as a semiclassical approximation to a quantum theory of gravity in the sense that the
existence of a minimal length is a presupposition for such a theory and not a consequence.
Therefore it would be very interesting to consider the complete quantum theory of the presented generalized theory of gravity.
The modifications of electrodynamics and gravity as they are investigated in this paper are in principle in accordance with the
formulation of these gauge field theories on noncommutative space-time by using the star product and Seiberg Witten maps where
similar modifications with additional interaction terms are obtained
\cite{Calmet:2001na},\cite{Calmet:2003jv},\cite{Calmet:2005qm},\cite{Calmet:2006zy}.
This is to be expected, since the concept of noncommutative space-time, which also implies a minimal length, is closely related
to the generalized uncertainty principle, which also leads to noncommuting coordinates.\\ \noindent
$Acknowledgement$:
I would like to thank the Messer Stiftung for financial support.

\end{document}